\documentclass[a4paper, 10 pt, conference]{ieeeconf}      
\IEEEoverridecommandlockouts                              
\usepackage{multicol}
\usepackage{graphics} % for pdf, bitmapped graphics files
\usepackage{amsmath} % assumes amsmath package installed
\usepackage{amssymb}  % assumes amsmath package installed
\usepackage{color}
\usepackage{siunitx}
\usepackage[table]{xcolor}
\usepackage[affil-it]{authblk}
\usepackage[T1]{fontenc}
\usepackage{graphicx}
\usepackage{rotating,stackengine,scalerel}
\usepackage{stfloats}
\usepackage[compress]{cite}
\newcommand\wye{\scalerel*{\stackengine{-1pt}{%
  \rotatebox[origin=c]{30}{\rule{10pt}{.9pt}}\kern-1pt%
  \rotatebox[origin=c]{-30}{\rule{10pt}{1.3pt}}}{%
  \rule{.9pt}{10pt}}{O}{c}{F}{F}{S}}{\Delta}}
\DeclareSIUnit{\rpm}{rpm}

\begin{document}

\twocolumn[
\begin{@twocolumnfalse}

\title{\LARGE \bf Implementation of an Electric Drive Digital Twin in a Networked Test Lab Environment}

\author{Frank Liebich$^1$$^*$}
\author{Sebastian Pahnke$^1$$^*$}
\author{Mario Stamann$^2$} 
\author{Roberto Leidhold$^2$}
\author{Andreas Scholz$^1$}
\affil{$^1$Institute for Engineering of Products and Systems of the Otto-von-Guericke-Universität Magdeburg, Germany}
\affil{$^2$Institute of Electric Power Systems of the Otto-von-Guericke-Universität Magdeburg, Germany}

\maketitle

\end{@twocolumnfalse}
]

\renewcommand{\footnoterule}{
  \vspace*{-0.5em}
  \noindent\rule{0.4\columnwidth}{0.4pt}
  \vspace*{0.4em}
  }
  
\renewcommand\thefootnote{*}
\footnotetext{These authors contributed equally to this work.}
\renewcommand\thefootnote{\arabic{footnote}}
\vspace{0.5em}

\begin{abstract}

The concept of Digital Twin (DT) consists of a physical asset, a digital asset, and their bidirectional data exchange, differing the DT from concepts with lower level of integration. Availability of the bidirectional interconnection not only enables monitoring of system states but also allows automated control of the system. Leveraging the DT concept, this paper presents a practical implementation of an electric drive DT by means of a permanent magnet synchronous machine (PMSM) in a test lab environment. The approach aims to virtually sense the disturbance variable, here the acting load torque. By integrating an observer controller into the PMSM control loop, the DT enables compensation of the disturbance variable, which is not measurable in many applications. The results of the DT implementation recorded on the test bench demonstrate an effective disturbance compensation of the acting load torque by the observer controller. However, the results still show a deviation between the observed and measured load torques, indicating room for further refinement. This work serves as a first step towards further practical applications of the DT in the electric drive testing environment.

%Digital twins of electric drive systems enable the monitoring of system states and can be used to provide additional information about the overall system that would otherwise only be available via sensors. Modern embedded control systems for electric machines usually have high computing power, so that additional functions implemented in software do not require more technical effort. Online digital twins thus open up new possibilities for improving diagnostic capabilities, predictability, and many other aspects.
%For example the online-detection of faults in electrical machines at an early stage by observation of difficult or non measurable state variables like acceleration, flux and temperature without the use of additional sensors. This article demonstrates the practical implementation of a model-based digital twin of a permanent magnet synchronous machine (PMSM) in a network-compatible test bench environment and represents only the first step toward further practical applications of the digital twin in electric drive technology.

\end{abstract}

\section{Introduction}

The digital transformation of the automotive industry fundamentally redefines the way vehicles are developed, tested, and brought to the market. Several terms, such as Industry 4.0, Cyber-Physical System, or Digital Twin (DT), are frequently used in this context. In particular, the term DT is widely used in the automotive industry and is often understood incorrect or incomplete \cite{Kritzinger.2018}. 
\newline Due to the lack of a common DT definition and the diverse understanding of its characteristics \cite{Bitencourt.2025}, only few publications demonstrate the concept and its potential in practical use cases. Therefore, this paper aims to develop a specific application of a DT utilizing a permanent magnet synchronous machine (PMSM) in a test bench environment. The potential of DT for virtual sensing of non-directly measurable quantities is emphasized by applying an existing framework for creating a DT.
\section{State of the Art}

\subsection{DT Definition and Framework}
With the increasing integration of Cyber-Physical Systems and real-time data connectivity in industrial contexts, the DT has become a fundamental concept within Industry 4.0. The concept of DT was first proposed in 2002 by Michael Grieves in the context of Product Lifecycle Management \cite{Grieves.2016}. Initially presented and not yet named explicitly, the term DT was coined by John Vickers in 2010. Since then, the concept of DT has gained significant attention in both academic research and industrial applications \cite{vanBossuyt.2025}. However, despite popularity, until now there has been a lack of a universally agreed definition. In fact, there are many different definitions in various industries, mainly varying in terms of characterization and application fields \cite{VanDerHorn.2021}. 
\newline According to the common literature \cite{Jones.2020,Hu.2025,Tao.2018,Melesse.2021,AboElHassan.2023}, a DT contains a virtual representation of a physical asset, system, or process, which reflects and predicts the current or future state, condition, or behavior of the physical asset \cite{Hu.2025}. It is continuously synchronized with its real-world counterpart through real-time data exchange, enabling comprehensive monitoring, analysis, prediction, decision making, and control of the physical entity \cite{Melesse.2021}. A defining characteristic of DT is the fully automated, real-time-capable, bidirectional data exchange between the physical and digital entity \cite{Grieves.2016}. The DT combines continuous sensor data measurement with feedback and provides control signals back to the real system in order to affect the system state. This automated bidirectional data exchange separates the concept of DT from other approaches with a lower level of integration, such as Digital Shadow and Digital Model \cite{Kritzinger.2018}. 
\newline Due to increasing complexity and to ensure interoperability, scalability, and real-time capabilities, a variety of frameworks for the creation of DTs have been proposed in recent years, mainly in the manufacturing context. These frameworks aim to provide a structured approach to model, connect, and manage the interaction between physical and digital entities. Some conceptual approaches primarily provide methodological frameworks for the development of DT, such as the Four Rs framework \cite{Osho.2022} or the Product Design Framework \cite{Tao.2019}. Other frameworks also offer a practical approach by proposing general architectures to build DTs, as shown in \cite{Andrade.2021} and \cite{Zhang.2021}, or adaptive DTs, as shown in \cite{Ogunsakin.2023}. Furthermore, the ISO 23247:2021 standard provides guidance for the development of DTs, especially in the manufacturing environment \cite{InternationalOrganizationforStandardization.}.
\newline As with the definition of DT, there is a lack of commonly accepted frameworks and publications demonstrating the low-level creation of DTs using existing approaches.

\subsection{Field Oriented Control of PMSM}
    \label{chap:Field_Oriented_Control_of_PMSM}

The field-oriented control system of a three-phase PMSM is a fundamental part of a drive system, enabling precise control of torque and rotational speed. Under this assumption, the electrical part of a PMSM can be described by two differential equations for the $d$- and $q$-components of the stator currents, as follows:

\begin{equation}
    \dfrac{di_d}{dt}=\dfrac{1}{L_W} \cdot \left( u_d - R_W \cdot i_d - L_W \cdot \omega_e \cdot i_q \right)
    \label{equ:did_dt}
\end{equation}
\begin{equation}
    \dfrac{di_q}{dt}=\dfrac{1}{L_W} \cdot \left( u_q - R_W \cdot i_q - L_W \cdot \omega_e \cdot i_d - \omega_e \cdot \Psi_{\mathrm{PM}}\right)
    \label{equ:diq_dt}
\end{equation}
Constant parameters are the stator winding resistance $R_W$, the stator winding inductance $L_W$ and the flux linkage $\Psi_{\mathrm{PM}}$ of the permanent magnets. The electrical frequency $\omega_e$ and the currents $i_d$ and $i_q$ are time-dependent measurable state variables in a rotor-oriented reference frame. In this reference frame, mechanical torque $M_M$ is generated directly by the current state variable $i_q$ and the constant flux linkage $\Psi_{\mathrm{PM}}$ according to Eqn. \ref{equ:torque}.

\begin{equation}
    M_M=\dfrac{3}{2} \cdot z_P \cdot i_q \cdot \Psi_{\mathrm{PM}}
    \label{equ:torque}
\end{equation}
The $i_d$ current is controlled to zero, which means that in this study no reluctance torque is used.
The mechanical part of the electric drive system is described in Eqn. \ref{equ:mechanical_dgl} and consists of the moment of inertia $J$, the acting load torque $M_L$ and the speed-dependent friction parameter $k_{\omega}$.

\begin{equation}
    \dfrac{d\omega_{m}}{dt}=\dfrac{1}{J} \cdot \left( M_M-M_L-k_{\omega} \cdot \omega_{m} \right)
    \label{equ:mechanical_dgl}
\end{equation}
Using the number of pole pairs $z_P$, the mechanical rotational speed of the shaft $\omega_{m}$ is calculated by $\omega_e=\omega_{m} \cdot z_P$. \\
Two current controllers apply the stator voltages $u_d$ and $u_q$ to control the mechanical torque $M_M$ to the desired reference value. A superimposed speed controller ensures that the steady state shaft speed is independent of load torque or follows the reference value as fast as possible. This requires direct measurement of the phase currents and the rotor speed or position, along with a real-time control system to implement field-oriented control of the inverter.
\section{Development of the DT}

\subsection{DUT and Test Bench}
    \label{chap:DUT_and_Test_Bench}

\begin{table}[t]
    \centering
    \caption{Determined system parameters of DUT.}
    \begin{tabular}{|c|c|c|c|}
        \hline
        \textbf{Parameter} & \textbf{Symbol} & \textbf{Value} & \textbf{Unit} \\ \hline
        Rated speed  & $n_N$ & 1000 & $\SI{}{\rpm}$ \\ \hline
        Maximum speed & $n_{\mathrm{max}}$ & 2000 & $\SI{}{\rpm}$ \\ \hline
        Rated torque & $T_N$ & 600 & $\SI{}{\N}\SI{}{\m}$ \\ \hline
        Peak torque & $T_{\mathrm{max}}$ & 800 & $\SI{}{\N}\SI{}{\m}$ \\ \hline
        Rated power & $P_N$ & 60 & $\SI{}{\kilo\watt}$ \\ \hline
        DC link voltage & $U_{\mathrm{DC}}$ & 560 & $\SI{}{\V}$ \\ \hline
        Rated current & $I_N$ & 210 & $\SI{}{\A}$ \\ \hline
        Number of pole pairs & $z_P$ & 5 & $-$ \\ \hline
        Winding resistance per phase & $R_W$ & 50 & $\SI{}{\m}\SI{}{\Omega}$ \\ \hline
        Inductance per phase & $L_W$ & 0.95 & $\SI{}{\mH}$ \\ \hline
        Flux linkage & $\Psi_{\mathrm{PM}}$ & 0.38 & $\SI{}{\V}\SI{}{\s}$ \\ \hline
        Motor inertia & $J$ & 2.1966 & $\SI{}{\kg}/\SI{}{\m^2}$ \\ \hline
    \end{tabular}
    \label{tab:DUTparameters}
\end{table}

\begin{figure}[!b]
    \includegraphics[width=\linewidth]{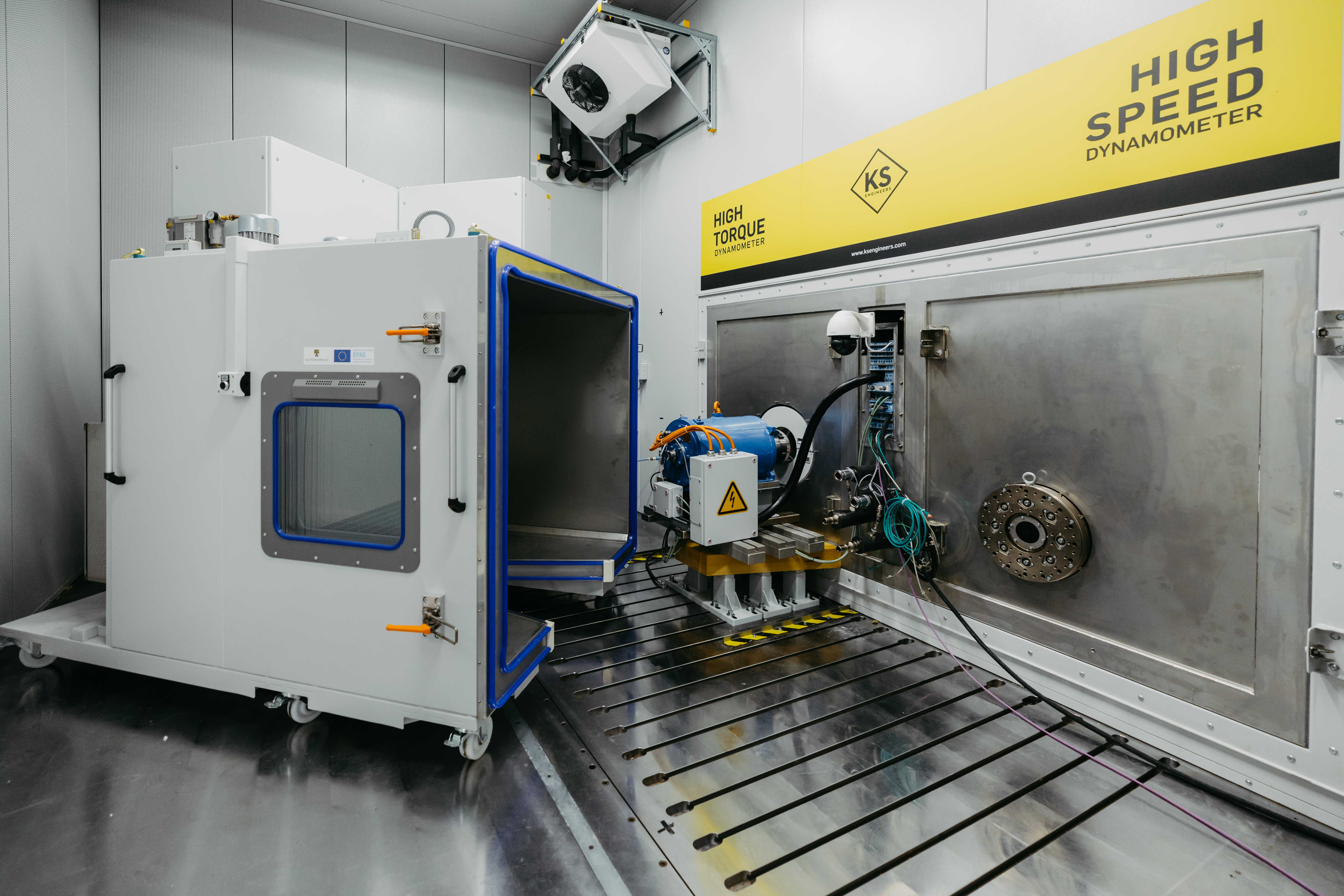}
    \caption{DUT mounted on the high-torque test bench at CMD in Barleben, mechanically coupled to the load machine (not visible in the figure).}
    \label{fig:DUT&TB}
\end{figure}

\begin{figure*}[!b]
\centering
    \includegraphics[width=\textwidth]{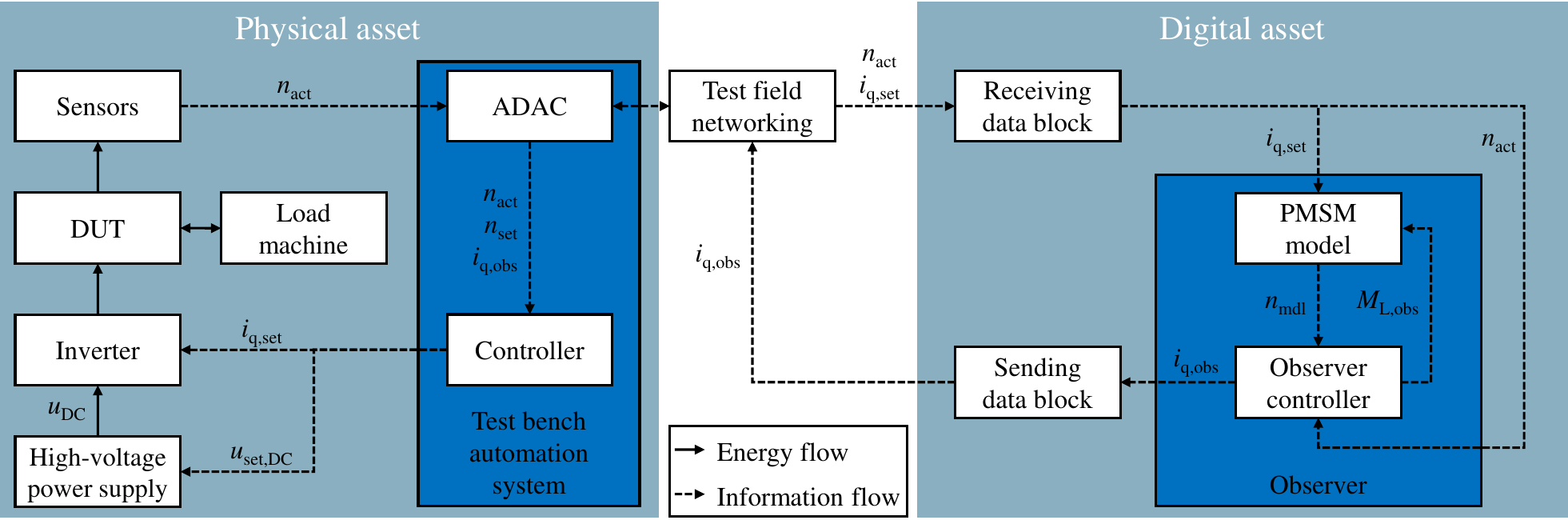}
    \caption{DT Framework adapted from \cite{Andrade.2021}, schematically representing the physical and digital assets and their bidirectional signal flow.}
    \label{fig:DT_Framework}
\end{figure*}

In order to develop a practical DT application, this contribution uses a PMSM featuring permanent magnets buried in the rotor as the device under test (DUT).
Table \ref{tab:DUTparameters} summarizes the system parameters determined for the physical DUT.
The DUT is implemented on a dedicated test bench for electric machines within the Center for Method Development (CMD) in Barleben. As depicted in Fig. \ref{fig:DUT&TB}, the test bench comprises two test slots. One test slot, in which the DUT is mounted, is designed to test electric machines on high-torque applications up to $\SI{3500}{\N}\SI{}{\m}$. The other test slot is used to test electric machines on high-speed applications up to $\SI{23000}{\rpm}$. The mechanical power output of the DUT can be absorbed and manipulated by a load machine on the counter side of the DUT. Both electric machines in the high-torque slot are mechanically coupled by a torque transducer with a nominal torque of 5 $\SI{}{kNm}$.
\newline The load machine and the DUT are operated using a real-time measurement and control system (called ADAC), comprehensive measurement technology, and a host automation system. Furthermore, the test bench includes a DUT conditioning system (air and fluids) and an acoustic chamber to measure the noise, vibration, and harshness (NVH) emissions of the DUT.
\newline At the CMD, the electric machine test bench is integrated into a networked test environment. This environment also contains several other test facilities, such as a complete vehicle test bench, a battery test bench, and a fuel cell test bench. A hardware-in-the-loop (HiL) simulator is integrated into the test environment to enable advanced testing options, such as the implementation of DTs. Using an ethernet-based network concept, the test benches are interconnected in real time, with signal sampling rates of up to 10 kHz. This structure not only supports networked tests ranging from the component level to the complete system level, but also enables virtual integration and virtual testing within the test environment.

\subsection{Modelling of DT}
    \label{chap:Modelling_of_DT}

The DT is developed using a framework based on the concept initially proposed by Andrade et al. \cite{Andrade.2021} for DTs in industrial processes. The original framework is characterized by a clear separation between the physical and the digital assets, which are connected via a standardized data exchange server. In the present work, this architecture is adapted to the specific test setup, as illustrated in Fig. \ref{fig:DT_Framework}.
\newline 
The physical asset of the DT contains the DUT, which is mounted on the test bench and mechanically coupled to the load machine. The DUT is electrically connected to a universal inverter with an integrated control system, which is powered by a high-voltage DC supply unit and operated by the test bench automation system. To operate the DUT, the DC-link voltage $u_{\mathrm{DC}}$ for the inverter is set by the automation system, while field-oriented current control is selected as the operating mode of the inverter control system. A proportional (P) speed controller implemented on the ADAC determines the torque-generating reference current $i_{\mathrm{q,set}}$, based on the deviation between the reference rotational speed $n_{\mathrm{set}}$ and the actual rotational speed $n_{\mathrm{act}}$ measured by the PMSM encoder.
Figure \ref{fig:Disturbance_Observer} shows the signal flow diagram of the DT containing the DUT and a model-based disturbance observer for the load torque $M_L$. The gain $k_p$ of the speed controller is optimized according to the magnitude optimum criterion, as follows:

\begin{figure}[t]
    \centering
    \includegraphics[width=\linewidth]{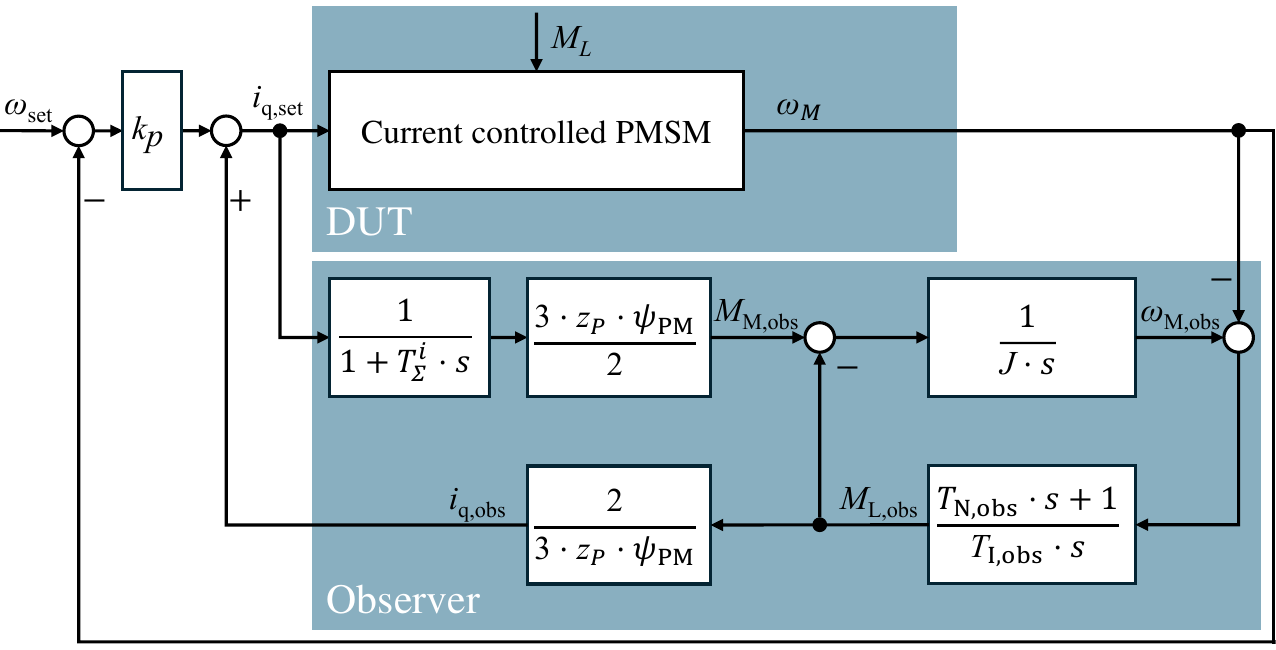}
    \caption{Signal flow diagram of the DT consisting of DUT and disturbance observer.}
    \label{fig:Disturbance_Observer}
\end{figure}

\begin{equation}
    k_{p}=\dfrac{1}{2 \cdot k_{s} \cdot T_{\Sigma}^i}
    \label{equ:kp_optimum_magnitude_criterion}
\end{equation}
Here, $T_{\Sigma}^i$ is defined as the sum of all small time constants of the speed control loop, including the dynamic behavior of the current control loop of the DUT. The gain $k_{s}$ follows from the ideal plant transfer function defined in Eqn. \ref{equ:torque} and \ref{equ:mechanical_dgl} of the mechanical system, assuming $k_{\omega}=0$. This leads to the following expression for the controller gain $k_p$:

\begin{equation}
k_{p}= \frac{J}{3 \cdot n_P \cdot \Psi_{\mathrm{PM}} \cdot T_{\Sigma}^i} 
\end{equation}
\label{equ:system_gain_kp} 
\newline
The digital asset comprises the load torque observer, including a PMSM simulation model and a proportional-integral (PI) observer controller. After implementation on the HiL simulator, the observer is connected to dedicated data receiving functions. These functions transfer measurement and control data from the test bench automation system to the observer controller and the PMSM simulation model. Additionally, the observer feedback is transmitted back to the test bench automation system through dedicated data sending functions. The communication of the physical and digital assets is handled via the test field network using ADAC-Link, a proprietary packet-based protocol over ethernet that enables real-time communication within the test lab environment. 
\newline
In order to detect the load torque acting on the DUT, the observer leverages the deviation between the measured rotational speed of the DUT $n_{\mathrm{act}}$ and the rotational speed obtained by the PMSM model $n_{\mathrm{mdl}}$. For this purpose, the PMSM model is derived from the equation introduced in Section \ref{chap:Field_Oriented_Control_of_PMSM} and parameterized to emulate the static and dynamic behavior of the DUT.
\newline
When an external load torque acts on the DUT, its rotational speed decreases due to the limited ability of the P speed controller to compensate for disturbances. The simulation model, which receives the same input quantities as the DUT, remains unaffected by the load torque and maintains its state. The resulting rotational speed difference is captured and transmitted to the PI observer controller. Based on the rotational speed difference, the load torque acting on the DUT is estimated by the observer controller. Mathematically, the transfer function of the observer controller can be expressed in continuous-time form as follows:

\begin{equation}
G_{\mathrm{PI}}(s)= \dfrac{T_{\mathrm{N,obs}} \cdot s +1}{T_{\mathrm{I,obs}} \cdot s} = k_{\mathrm{p,obs}} + \frac{k_{\mathrm{i,obs}}}{s}
    \label{equ:Gpi(s)}
\end{equation}
The proportional gain $k_{\mathrm{p,obs}}$ and the integral gain $k_{\mathrm{i,obs}}$ result from the optimized disturbance compensation with $T_{\mathrm{N,obs}}=T_{\Sigma}^i$ and $T_{\mathrm{I,obs}}={T_\Sigma^i}^2 \cdot J^{-1}$ as shown in Eqn. \ref{equ:Gpi(s)}. Equation \ref{equ:Kp_and_Ki_obs} describes the analytical calculation of the independent gain parameters of the PI controller.

\begin{equation}
k_{\mathrm{p,obs}} = \frac{J}{T_{\Sigma}^i}, \quad k_{\mathrm{i,obs}} = \frac{J}{{T_{\Sigma}^i}^2} 
    \label{equ:Kp_and_Ki_obs} 
\end{equation}
As shown in Fig. \ref{fig:DT_Framework} and \ref{fig:Disturbance_Observer}, the observer controller feeds back the observed load torque $M_{\mathrm{L,obs}}$ to the PMSM model as an additional resistive torque, aligning the model's rotational speed with that of the DUT. In steady state, this observed torque is intended to equal the real load torque $M_L$, assuming that the error between $\omega_M$ and $\omega_{\mathrm{M,obs}}$ is zero and the model parameters $T_\Sigma^i$, $\Psi_{\mathrm{PM}}$, and $J$ are exactly known.
The observed load torque can then be converted into a corresponding current value $i_{\mathrm{q,obs}}$ according to Eqn. \ref{equ:torque} and transmitted to the physical asset, where it is added to the reference current $i_{\mathrm{q,set}}$ (see Fig. \ref{fig:DT_Framework}).
%Furthermore, it is possible to convert the observed load torque into a corresponding current value $i_{\mathrm{q,obs}}$ according to Eqn. \ref{equ:torque} and transmit it to the physical asset, where it is added to the reference current $i_{\mathrm{q,set}}$ (see Fig. \ref{fig:DT_Framework}). 
This allows the DT to compensate for the load torque directly and supports the limited ability of the P speed controller to handle disturbances and adjust the rotational speed difference of the DUT.

% -> PMSM-Modell 
%  -> Auslegung der Systemparameter:
%     Bestimmung J: symm. Hoch-/Tieflauf unter annahme konst. Widerstandsmomente 
%     Bestimmung kw: empirische Bestimmung durch Drehzahldifferenz zwischen DUT und Beobachtermodell infolge unterschiedlicher Widerstände
%     Beschreibung Prüfling (Datenblatt, Systemparameter)
\section{Evaluation of DT}

In the following sections, the DT is evaluated by performing test scenarios on the test bench setup in order to compare the physically measured data with the data simulated in the digital asset. Within this approach, the digital asset is evaluated by determining the accuracy of the PMSM model and by analyzing the behavior of the system with and without the observer controller activated. The evaluation is performed using two measurement scenarios. 
\newline
Scenario A applies a load torque change while maintaining a constant rotational speed set point (see Fig. \ref{fig:plotLastsprung}). Scenario B refers to a realistic vehicle speed profile derived from the Worldwide harmonized Light-duty vehicles Test Cycle (WLTC) class 3 (see Fig. \ref{fig:plotWLTC}).
The graphs of both scenarios are divided into two sections. The upper sections show a comparison of the DUT’s measured rotational speed with the target value and the model's rotational speed. The lower sections display the measured load torque compared to the set point and the observed torque.
\newline
In both scenarios, driving resistance is determined using a vehicle longitudinal dynamics model that neglects slip, as described in \cite{Schutz.2016}. The model calculates the driving resistance based on driving kinematics and vehicle characteristics. For this purpose, a notional vehicle is defined, whose parameters are summarized in Tab. \ref{tab:vehicleParameters}.

\begin{table}[b]
    \centering
    \caption{Parameters of notional vehicle.}
    \begin{tabular}{|c|c|c|c|}
        \hline
        \textbf{Parameter} & \textbf{Symbol} & \textbf{Value} & \textbf{Unit} \\ \hline
        Drivetrain efficiency & $\eta_T$ & 0.95 & $-$ \\ \hline
        Rolling resistance coefficient & $f_R$ & 0.01 & $-$ \\ \hline
        Gravitational acceleration & $g$ & 9.81 & $\SI{}{\m}/\SI{}{\s^2}$ \\ \hline
        Vehicle mass & $m_F$ & 1000 & $\SI{}{\kg}$ \\ \hline
        Aerodynamic lift or downforce & $F_A$ & 0 & $\SI{}{\N}$ \\ \hline
        Incline & $q$ & 0 & $\SI{}{\%}$ \\ \hline
        Mass factor of rotating masses & $e$ & 1.1 & $-$ \\ \hline
        Aerodynamic drag coefficient & $c_w$ & 0.25 & $-$ \\ \hline
        Frontal area & $A_x$ & 2.2 & $\SI{}{\m^2}$ \\ \hline
        Air density & $\rho_L$ & 1.197 & $\SI{}{\kg}/\SI{}{\m^3}$ \\ \hline
        Wheel radius & $r_{\mathrm{wheel}}$ & 0.25 & $\SI{}{\m}$ \\ \hline
        Gear ratio & $i$ & 1:1.6 & $-$ \\ \hline
    \end{tabular}
    \label{tab:vehicleParameters}
\end{table}

\subsection{Evaluation of PMSM Model}

\begin{figure}[!t]
    \centering
    \includegraphics[width=9cm]{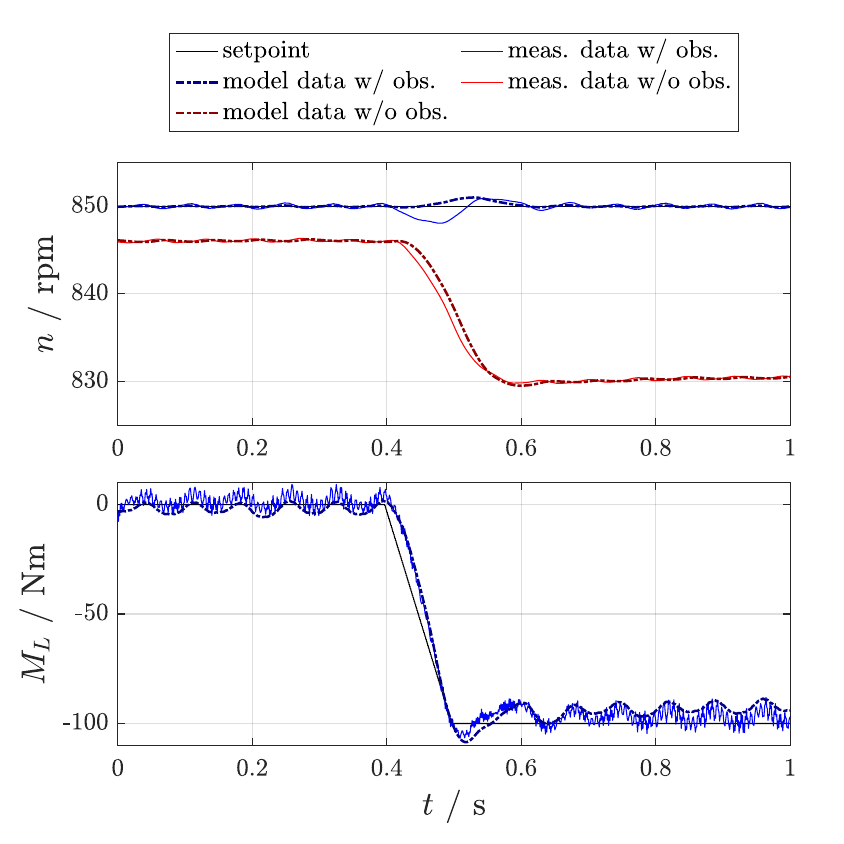}
    \caption{Comparison of measured and predicted model data by means of rotational speed (upper graph) and load torque (lower graph) specifying a torque load change at constant speed setpoint.}
    \label{fig:plotLastsprung}
\end{figure}

The aim of creating and parameterising the PMSM model is to optimally emulate the physical behavior of the real electric machine. Therefore, the deviation between the data simulated in the PMSM model and the measured data is analyzed in both scenarios. 
In scenario A (Fig. \ref{fig:plotLastsprung}), the mean absolute error (MAE) between the measured and simulated rotational speed of the PMSM is $\SI{0.3}{\rpm}$, regardless of the activation of the observer controller, corresponding to a mean absolute percentage error (MAPE) of $\SI{0.04}{\%}$. Similarly, the MAE in scenario B (Fig. \ref{fig:plotWLTC}) amounts to $\SI{1.5}{\rpm}$, equivalent to a MAPE of $\SI{0.41}{\%}$. This evaluation shows that the PMSM model within the digital asset emulates the actual measured rotational speed accurately under the investigated operating conditions. Analyzing scenario A, it is visible that rapid transient conditions, such as the rapid change in load torque, result in a temporary deviation of the model from the real system. However, the model can predict the actual rotational speed precisely under mainly stationary conditions.

\subsection{Evaluation of Load Torque Observer}

\begin{figure}[!t]
    \centering
    \includegraphics[width=9cm]{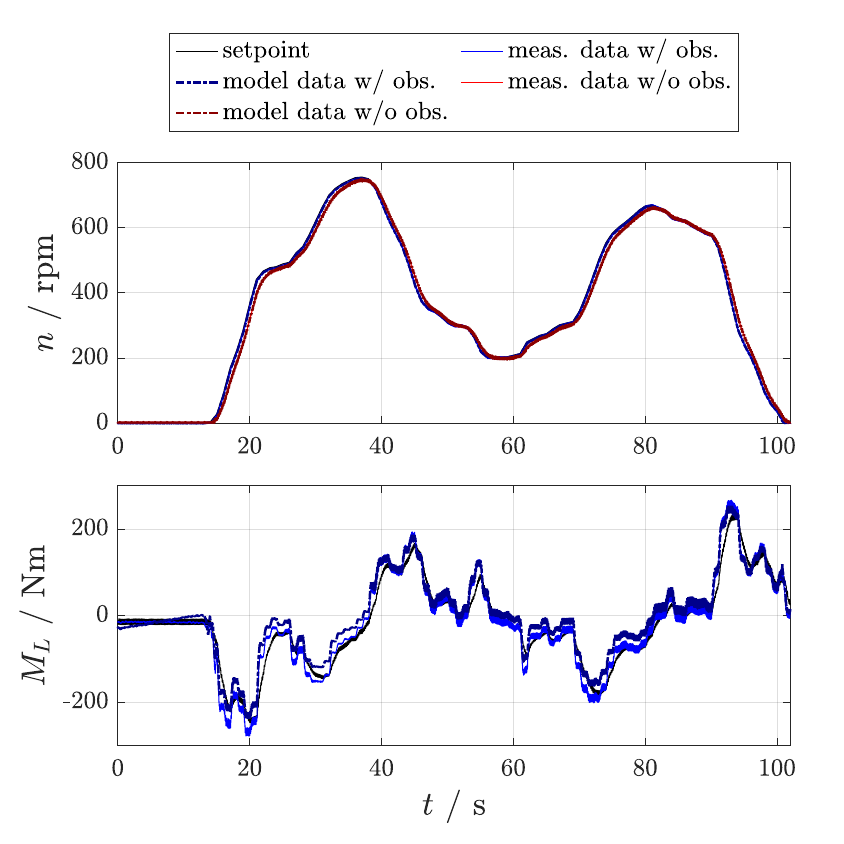}
    \caption{Comparison of measured and predicted model data by means of rotational speed (upper graph) and load torque (lower graph) specifying a speed profile according to WLTC class 3 and corresponding driving resistances.}
    \label{fig:plotWLTC}
\end{figure}

The load torque observer acts as a virtual sensor within the digital asset of the DT, determining the load torque affecting the DUT. Based on the longitudinal dynamics model used, the effective load torque primarily depends on the vehicle's driving kinematics. 
Scenario A (Fig. \ref{fig:plotLastsprung}) describes a constant rotational speed set point of 850 $\SI{}{\rpm}$, equivalent to a driving speed of 50 $\SI{}{\km}/\SI{}{h}$ based on the vehicle parameters used. The applied load torque of 100 $\SI{}{Nm}$ is proportional to the driving resistance of the estimated vehicle speed. The qualitative courses of the rotational speed curves in scenario A show that the observer controller's disturbance compensation works effectively and the rotational speed of the electric machine is kept constant regardless of the acting load torque. In contrast, mere control through a P speed controller does not compensate for friction torque, as is noticeable by the rotational speed deviation at the beginning of the measurement in Fig. \ref{fig:plotLastsprung}. Furthermore, additional load torque also contributes to a loss of rotational and vehicle speed without the use of the PI observer controller.
\newline
Scenario B (Fig. \ref{fig:plotWLTC}) shows a practical application of the DT using a real driving cycle. It contains the rotational speed profile resulting from the first 100 seconds of WLTC class 3. Utilizing the longitudinal dynamics model, the translational speed of the vehicle is converted to the rotational speed of the electric machine. In addition, the driving resistances caused by the vehicle kinematics are calculated and applied to the DUT by setting the load torque according to the driving condition.
\newline
The analysis of the rotational speed curves in both scenarios emphasizes the effect of an observer controller in contrast to a simple P speed controller. In scenario B, the calculated MAE of the measured rotational speed to the target value is $\SI{3.5}{\rpm}$ when the observer controller is activated. Deactivating the controller increases the MAE to 14.1 $\SI{}{\rpm}$, which is also reflected in a slow-responding control behavior of the rotational speed curve.
Similarly to scenario B, the discrepancy between the rotational speed curves in scenario A further underlines the performance of the observer controller.
\newline
As with the rotational speed, the qualitative course of the observed load torque correlates with the target and the measured torque in both scenarios. 
For evaluation purposes, only the load torque recorded with the observer controller activated is analyzed and illustrated, since the deviations between observed and measured load torque are identical to those obtained when the controller is deactivated. However, there are significant deviations between the load torque determined by the observer and the load torque actually measured by the test bench. In absolute terms, the MAE between the observed and the actual measured load torque is 20.8 $\SI{}{Nm}$ in scenario A and 10.4 $\SI{}{Nm}$ in scenario B. These significant deviations are also recognizable in the displayed measurement curves and indicate inaccurate observation of the actual load torque.

\section{Conclusion and Future Research}

This paper presents a practical application demonstrating the concept and potential of a DT for virtual sensing of non-directly measurable quantities. It shows how the DT concept, which has been mostly discussed theoretically in recent research, is applied to a practical automotive use case of an electric drive within a networked test lab environment. Adapting an existing framework, the creation and implementation of a DT is explained, considering the clear distinction between physical and digital assets while simultaneously enabling bidirectional data exchange.
\newline
This paper focuses on the potential of the DT to determine the load torque acting on a notional vehicle and consequently on the PMSM. Observing the driving resistance through the DT offers several advantages. Tests on a dedicated test bench can be simplified and carried out more cost-effectively, as the DT eliminates the need for a torque sensor. The approach can also be extended to road tests, where only the rotational speed of the PMSM must be measured to obtain information about the driving resistance.
In addition to testing, the PMSM DT can contribute to optimizing vehicle performance and battery range during regular driving operation by providing continuous data on driving resistances.
\newline
While these benefits highlight the potential of the DT approach, several challenges were encountered during its development. Measurement noise introduced by the test setup — including components such as the PMSM encoder and torque transducer, as well as effects like resonance phenomena and cogging torques — affected the accuracy of the observed quantities. The system identification of the DUT and the associated controller design were also influenced by the experimental setup and necessitated iterative refinement. In addition, a stable, real-time-capable communication infrastructure had to be established to ensure reliable data exchange between the physical and digital assets. Finally, the realization of the DT required an adaptable implementation-oriented framework that could be applied directly to the given use case.
\newline
Despite addressing these challenges, the DT evaluation performed in this paper reveals a deviation between the observed and measured load torque. Although the load torque compensation of the observer controller is effective, the observed torque values differ from the measured ones. Insufficient controller design or inaccurate measuring equipment may be potential reasons for the lack of fidelity of the load torque observer. In addition, general model uncertainties in the PMSM model also affect the estimated state variables under transient and steady-state conditions. Therefore, further research is needed to optimize the performance of the PMSM model and the observer controller, as well as to reduce deviations from sensor data. The observer will be extended in subsequent research in order to emulate the driving resistance more precisely and to detect mechanical or electrical faults, such as inter-turn short circuits or demagnetization of the stator magnets. For this purpose, the aim is to extend the DT of the PMSM by adding further physical domains to the model within the digital asset, thereby creating a multi-domain model. This can also be accomplished by exploring alternative modeling strategies for the digital asset, such as data-driven models or hybrid approaches. 
\newline
Beyond its application to the PMSM presented in this initial work, the DT concept will be applied to other components of the vehicle powertrain in future research, for example, the inverter or the traction battery. Using the capabilities of the specific component test benches and the fully networked test lab environment at the CMD, the DT concept will be investigated at the component level, subsystem level and complete system level. In this way, it will be possible to test all powertrain components both individually and jointly in an integrated powertrain using a co-simulation model. Therefore, emulating a complete vehicle within the CMD environment, leveraging the dedicated test benches, and applying the DT concept to the powertrain components are elements of subsequent research work.
\newline
The data analysis aspect of the current DT will also be extended using standardized interface concepts to enable the DT for online monitoring of system states and online parameter estimation. Using artificial intelligence or machine learning methods, an adaptive DT will be developed that is capable of deriving system parameters from measurement data during operation based on a baseline model. This will also contribute to a more accurate emulation of the DUT's behavior.

\bibliographystyle{ieeetr}
\bibliography{literature/bibliography}

\end{document}